\documentstyle[preprint,aps]{revtex}

\def\be{\begin{equation}}
\def\ee{\end{equation}}
\def\bea{\begin{eqnarray}}
\def\eea{\end{eqnarray}}

\def\bt{\begin{table}}
\def\et{\end{table}}
\def\ra{\rightarrow}
\def\ln{\left<}
\def\rn{\right>}
\def\lln{\left(}
\def\rln{\right)}
\def\lq{\Lambda_{QCD}}
\def\bl{\bar \Lambda}

\def\lbb{\Lambda_b}
\def\lq{\Lambda_{QCD}}
\def\bl{\bar \Lambda}

\def\ra{\rightarrow}

\def\om{\omega}
\def\qq{\ln \bar q q \rn}
\def\al{\alpha}

\begin{document}

\title{On the short distance nonperturbative corrections in
heavy quark expansion}
\author{S. Arunagiri}
\address{Department of Nuclear Physics, University of Madras\\
Guindy Campus, Chennai 600 025, Tamil Nadu, INDIA}
\maketitle
\begin{abstract}
We estimate the renormalon corrections to the inclusive decay 
rate of heavy hadrons. We assume the gluon mass term $\lambda^2
\gg \lq^2$ to imitate the short distance nonperturbative effects.
It is found that the the inclusive decay rates are corrected by
an amount of about 0.2 $\Gamma_0$ for $B$ meson and 0.21 $\Gamma_0$
for baryon. This is significant in view of the discrepancy of
lifetimes of $B$ and $\lbb$.
\end{abstract}
\vskip0.5cm
 
The divergence of perturbation theory at large order brings in
an ambiguity to physical quantities specified at short distances.
According to the present understanding, the ambiguity is given
by a class of renormalon diagrams which are chain of
$n$-loops in a gluon line. The phenomenon is deeply connected with
the operator product expansion (OPE). The perturbative part of the OPE
receives the renormalons corrections \cite{zak}.
Since in the OPE the first power-suppressed nonperturbative term is absent
and the renormalons corrections constitute short distance
nonperturbative effect, it is much meaningful beyond mere
large order corrections.

The phenomenology of the power corrections is significant
for the heavy quark expansion (HQE) to describe the
inclusive decays of heavy hadrons by an expansion in the
inverse powers of the heavy quark mass, $m_Q$.
As the inclusive decay rate of heavy hadrons scales like
fifth power of the heavy quark mass,
the power corrections arise due to momenta smaller than the
heavy quark mass. However, these IR renormalons would,
being nonperturbative effect, have greater influence in
the HQE prediction of quantities of interest.
These short distance nonperturbative effects
can be sought for explaining the smaller lifetime of $\Lambda_b$.
We should note that these power corrections to heavy quark decay rate
represents the breakdown of quark-hadron duality. Therefore, it may shed
light on the working of the assumption of quark hadron duality in
the heavy quark expansion.

In this note, we present a study on 
the renormalons corrections considering the
heavy-light correlator in the QCD sum rules approach, assuming that the
nonperturbative short distance corrections given by the gluon mass that is much
larger than the QCD scale. We carry out the analysis for both heavy meson
and heavy baryon. Our study shows that the short distance
nonperturbative corrections to the baryon and the meson
differ by a small amount which is significant for the smaller 
lifetime of the $\Lambda_b$.

Let us consider the
correlator of hadronic currents $J$:
\be
\Pi(Q^2) = i \int d^4x e^{iqx} \ln 0|T\{J(x)J(0)\}|0\rn
\ee
where $Q^2 = -q^2$. The standard OPE is expressed as
\be
\Pi(Q^2) \approx [~\mbox{parton model}](1+a_1 \al + a_2 \al^2 + ....)
+ O(1/Q^4)
\ee
where the power suppressed terms are quark and gluon operators.
The perturbative series in the above equation can be rewritten as
\be
D(\al) = 1 + a_0 \al + \sum_{n = 1}^\infty a_n \al^n \label{d}
\ee
where the term in the sum is considered to be the
nonperturbative short distance quantity.
It is studied by Chetyrkin {et al} \cite{chet} assuming that the short
distance tachyonic gluon mass, $\lambda^2$,
imitates the nonperturbative physics of the QCD. This, for the gluon
propagator, means:
\be
D_{\mu \nu}(k^2) = {\delta_{\mu \nu} \over {k^2}} \ra
\delta_{\mu \mu}\left( {1 \over {k^2}}+{\lambda^2 \over {k^4}}\right)
\ee
The nonperturbative short distance corrections are argued to be the
$1/Q^2$ correction in the OPE.

Let us consider the assumption of
the gluon mass $\lambda^2 \gg \lq^2$ which
is not necessarily to be tachyonic one.
The feature of the assumption can be seen
with the heavy quark potential
\be
V(r) = -{4\alpha(r) \over {3r}}+kr \label{v}
\ee
where $k \approx$ 0.2 GeV$^2$,
representing the string tension. It has been argued in
\cite{bal} that the linear term can be replaced by a term of order $r^2$.
It is equivalent to replace $k$ by a term describing the ultraviolet
region.
For the potential in (\ref{v}),
\be
k \ra constant \times \al \lambda^2
\ee
In replacing the coefficient of the term of $O(r)$ by $\lambda^2$,
we make it consistent by the renormalisation factor.
Thus the coefficient $\sigma(\lambda^2)$ is given by \cite{ani}:
\be
\sigma(\lambda^2) = \sigma(k^2)\left(\alpha(\lambda^2) \over
\alpha(k^2)\right)^{18/11} \label{rge}
\ee
Introduction of $\lambda^2$ brings in a small correction to the Coulombic term.
By use of (\ref{rge}), we specify the effect at both the ultraviolet
region
and the region characterised by the QCD scale.
Then, we rewrite (\ref{d}) as
\be
D(\al) = 1 + a_0 \al\left(1+{k^2 \over {\tau^2}}\right)
\label{dc}
\ee
where $\tau$ is some scale relevant to the problem and $k^2$ should be
read
from (\ref{rge}). We would apply this to
the heavy light correlator in heavy quark effective theory.

We should note that in the QCD sum rules approach, the scale involved
in is given by the Borel variable which is about 0.5 GeV. But
in the heavy quark expansion the relevant scale is
the heavy quark mass, greater than the
hadronic scale. Thus, there it turns out to be infrared renormalons
effects.
But, still it represents the short distance nonperturbative
property, by virtue of the gluon mass being as high as the hadronic
scale.

{\it Meson:} For the heavy light current, $J(x) = \bar Q(x) i\gamma_5
q(x)$,
the QCD sum rules is already known \cite{ben}:
\be
{\tilde f}_B^2 e^{-(\bl)} = {3 \over {\pi^2}}
\int_0^{\om_c} d \om \om^2 e^{-\om/\tau}D(\al) - \qq +
{1 \over {16 \tau^2}}\ln g \bar q \sigma G q \rn +...\label{hl}
\ee
where $\om_c$ is the duality interval, $\tau$ the Borel variable and
$D(\al)$ as defined in (\ref{d}), but of the form defined in (\ref{dc}).
It is, corresponding to the particular problem of heavy quarks, given as:
\be
D(\al)_B = 1 + {a_B \al}\left[1+{\lambda^2 \over {\tau^2}}
\left({\alpha(\lambda^2) \over {\alpha(\tau^2)}}\right)^{-18/11}\right]
\ee
where $a_B =   17/3+4\pi^2/9-4$log$(\om/\mu)$, with $\mu$
is chosen to be 1.3 GeV.

With the duality interval of about 1.2-1.4 GeV which is little smaller than
the
onset of QCD which corresponds to 2 GeV and $\bl \geq$ 0.6 GeV,
we get 
\be
\lambda^2 = 0.35 GeV^2. \label{l2m}
\ee

{\it Baryon}:
For the heavy baryon current
\be
j(x) = \epsilon^{abc}(\bar q_1(x)C \gamma_5 t q_2(x))Q(x)
\ee
where $C$ is charge conjugate matrix, $t$ the antisymmetric flavour matrix
and $a, b, c$ the colour indices, the QCD sum rules is given \cite{dai} by
\be
{1 \over 2} f_{\Lambda_b}^2 e^{\bl/\tau} =
{1 \over {20\pi^4}}\int_0^{\om_c} d \om \om^5
e^{-w/\tau}D(\al)_{\lbb}+
{6 \over {\pi^4}}E_G^4 \int_0^{\om_c} d \om e^{-\om/\tau}+
{6 \over {\pi^4}}E_Q^6e^{-m_0^2/8\tau^2}
\ee
where 
\be
D(\al)_{\lbb} = 1-{\al \over {4\pi}}a_{\lbb} \left(1+{\lambda^2 \over {\tau^2}}\right)
\ee
with $a_{\lbb} = r_1$log$(2\om/\mu)-r2)$. With
$f_{\Lambda_b}^2$ = $0.2 \times 10^{-3}$~GeV$^6$,
$\ln \bar q q \rn  = -0.24^3$ GeV$^3$, $\ln g \bar q \sigma G q \rn =
m_0^2 \ln \bar q q \rn$, $m_0^2 = 0.8$ GeV$^2$, $\ln \al GG \rn
= 0.04$ GeV$^4$ and $D(\al)_{\Lambda_b}$ is expressed in accordance with
power correction factor found in \cite{dai}.
As in the meson case, we obtain 
\be
\lambda^2 = 0.4 GeV^2.
\ee
Now we turn to the heavy quark expansion. The total decay
rate of a weakly decaying heavy hadron is, at the leading order, given by
\be
\Gamma(H_b) = \Gamma_0
\left[1-{\al \over \pi}\left( {2 \over 3}g(x) - \xi \right) \right] \label{rt}
\ee
where 
\be
\Gamma_0 = {G_f^2 |V_{KM}|^2 m_b^5 \over {192 \pi^3}} f(x)
\ee
As already mentioned, the power corrections are given by
the IR renormalons:
\be
\tilde a \al \lln 1+ {\lambda^2 \over {m_b^2}} \lln {\alpha(\lambda^2)
\over {\alpha(m_b^2)}} \rln ^{11/18} \rln
\ee
In (\ref{rt}), the factor $\xi$ corresponds to the IR renormalons
which corresponds to the square root of the $\lambda^2$ term in the
above equation. These corrections are estimated to be 0.1$\Gamma_0$
and 0.11$\Gamma_0$
for $B$ and $\Lambda_b$ respectively. This is significant in view
of the discrepancy between the lifetimes of $B$ and $\Lambda_b$
being 0.2 ps$^{-1}$ with $\Gamma(B)$ = 0.68 ps$^{-1}$ and
$\Gamma(\Lambda_b)$ = 0.85 ps$^{-1}$.

\end{document}